\documentclass{svproc}
\usepackage{url,doi,amsmath,amsfonts,amscd,amssymb,graphicx}

\newcommand{\R}{\mathbb{R}}

\newcommand{\C}{\mathbb{C}}

\newcommand{\CP}{\mathbb{C}P}

\def\bn{\boldsymbol{n}}
\def\be{\boldsymbol{e}}
\def\bF{\boldsymbol{F}}
\def\bA{\boldsymbol{A}}
\def\hA{\; \hat{\!\! \boldsymbol{A} }}
\def\bee{\begin{equation}}
\def\eee{\end{equation}}

\begin{document}
\mainmatter              
\title{Solvable Models of Magnetic Skyrmions}
\titlerunning{Magnetic Skyrmions}  
%
\author{Bernd  Schroers}
\authorrunning{Bernd Schroers} 
\institute{Department of Mathematics and Maxwell Institute for Mathematical Sciences, \\
Heriot-Watt University, Edinburgh EH3 9LW, UK \\
\email{b.j.schroers@hw.ac.uk}.
}

\maketitle              

\begin{abstract}
We give a succinct summary of the recently discovered solvable models of  magnetic skyrmions  in two dimensions, and of their general solutions. The models contain the standard Heisenberg term, the most general translation invariant Dzyaloshinskii-Moriya   (DM)  interaction term and, for each DM term, a particular combination of anisotropy and Zeeman potentials.  We
argue that simple mathematical features of the explicit solutions help understand  general qualitative properties of magnetic skyrmion configurations in more generic models.

\end{abstract}
\section{Introduction}
Magnetic skyrmions are topological solitons in two-dimensional field theories  for the magnetisation field $\bn$ of a magnetic material  \cite{BY,NT}. In the continuum version, the energy functional   consists of  the Dirichlet energy (quadratic in derivatives), a potential  which includes anisotropy terms and a Zeeman contribution 
(no derivatives), and the crucial Dzyaloshinskii-Moriya   (DM)  interaction term (linear in derivatives) \cite{Dzyaloshinskii,Moriya}. Such an energy functional has stationary configurations which are stable under Derrick scaling provided the DM term is negative for those configurations. 

Magnetic skyrmions have been the subject of intensive experimental and numerical studies in recent years because they combine interesting physics with potential technological applications in magnetic information storage  and manipulation \cite{FCS}. More recently,  rigorous analytical studies have established  conditions for the  existence of solutions as well as energy bounds in different topological sectors \cite{Melcher,DM}, and have clarified the interesting way in  which the relative energy of skyrmions and anti-skyrmions depends on the  DM term \cite{genDMI}.

\begin{figure}[ht]
\centering
\includegraphics[width=0.326\textwidth]{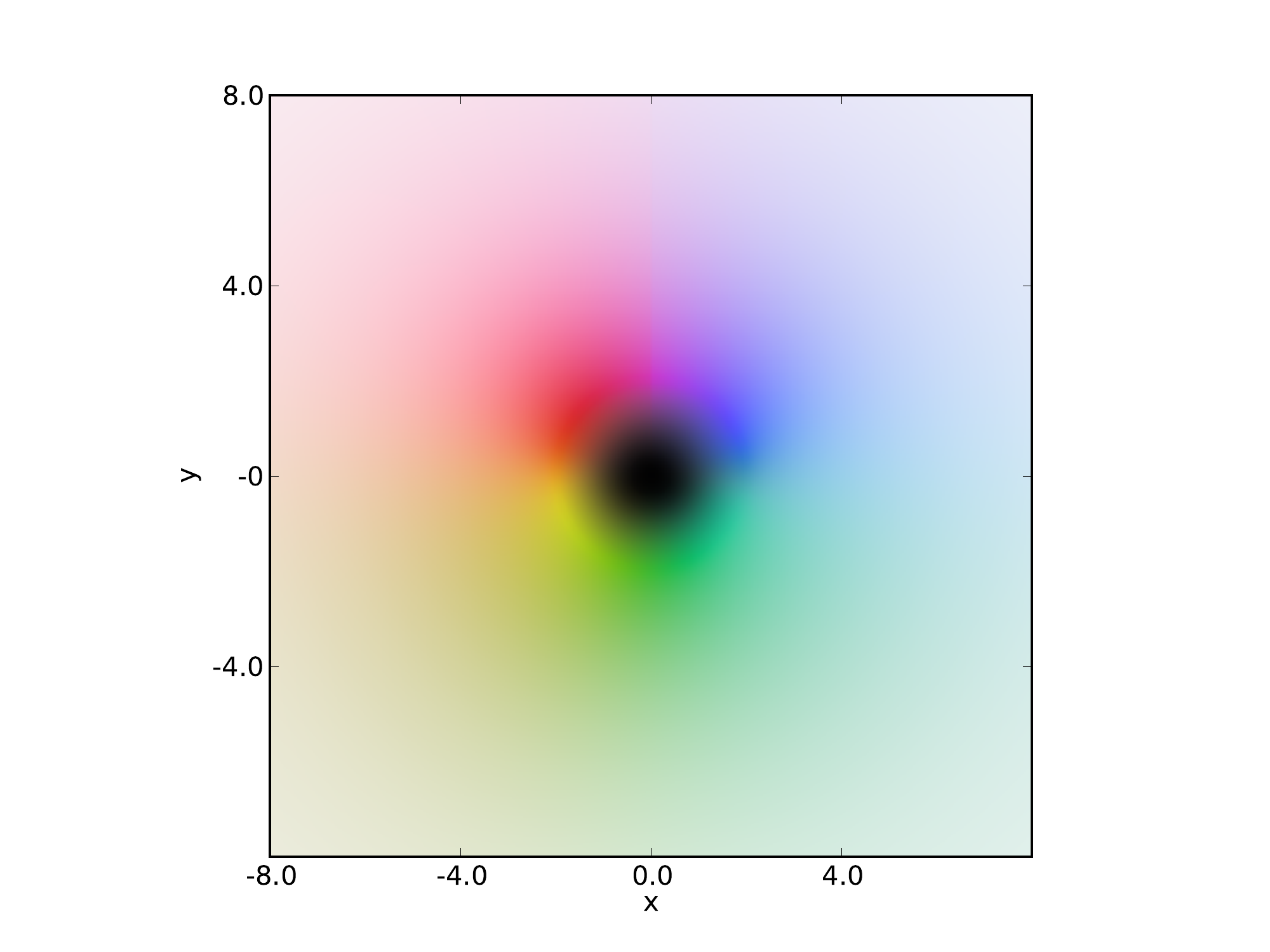}
\includegraphics[width=0.326\textwidth]{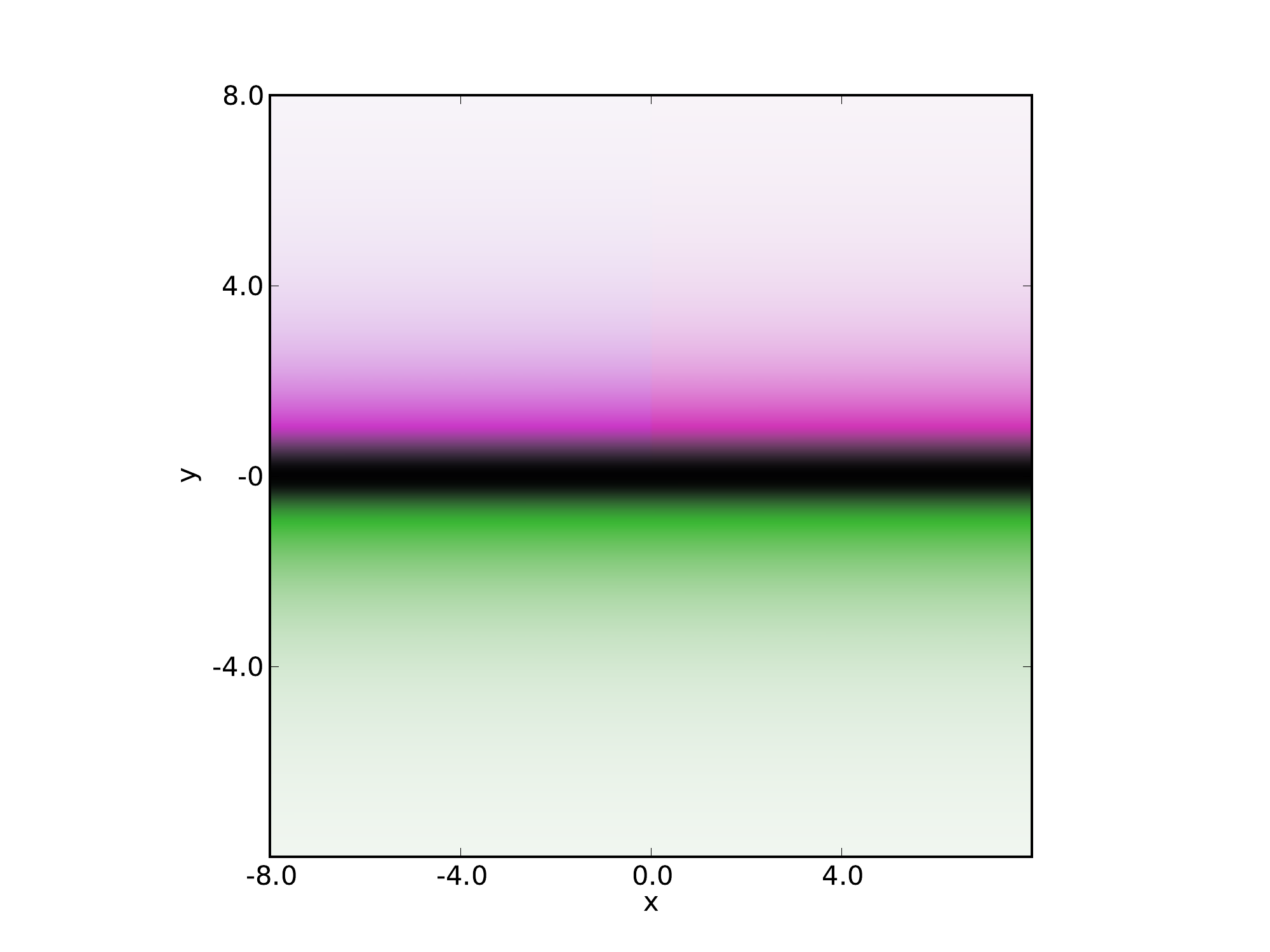}
\includegraphics[width=0.326\textwidth]{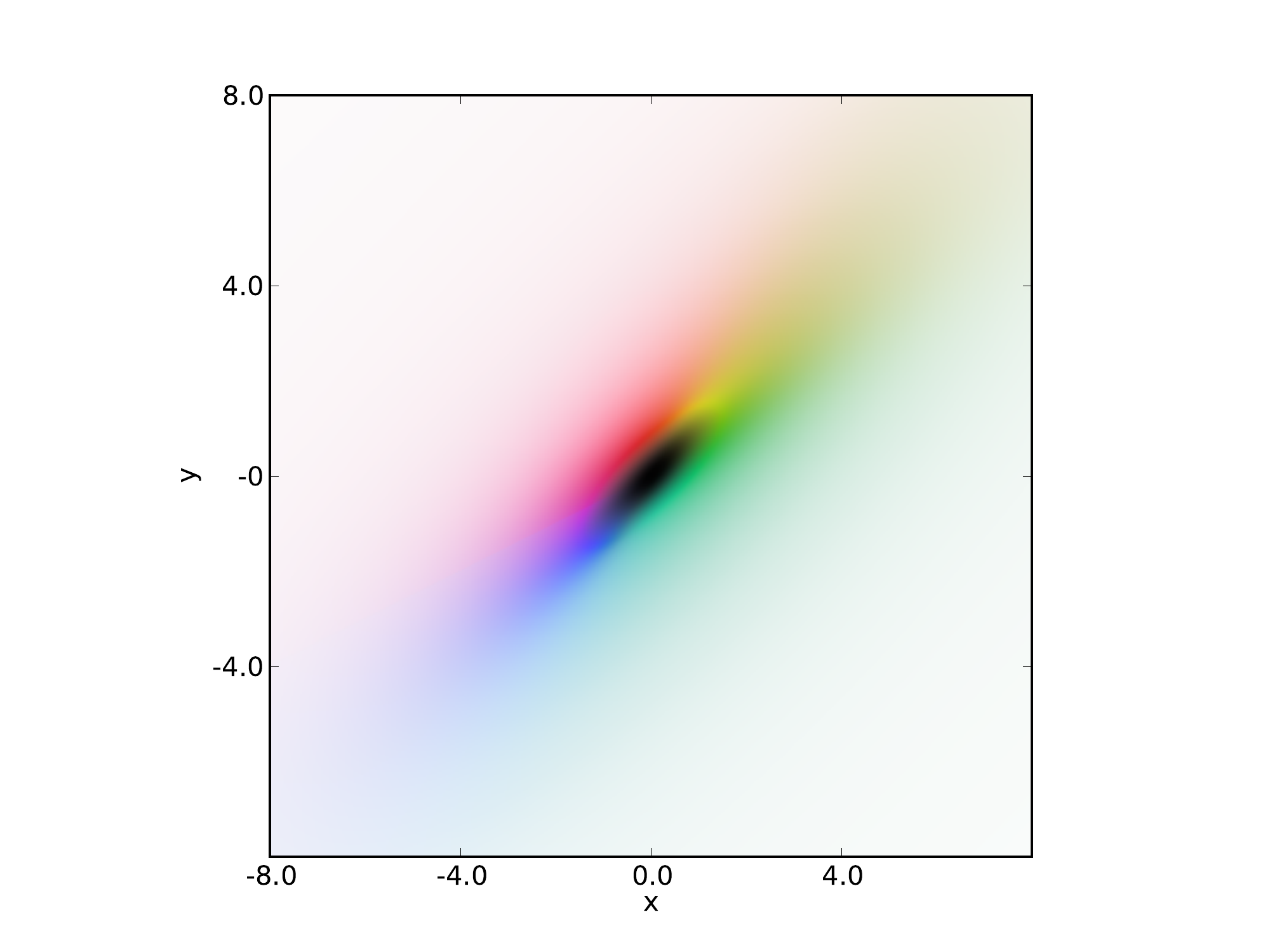} \\
 \includegraphics[width=0.326\textwidth]{2i_z_1}
   \includegraphics[width=0.326\textwidth]{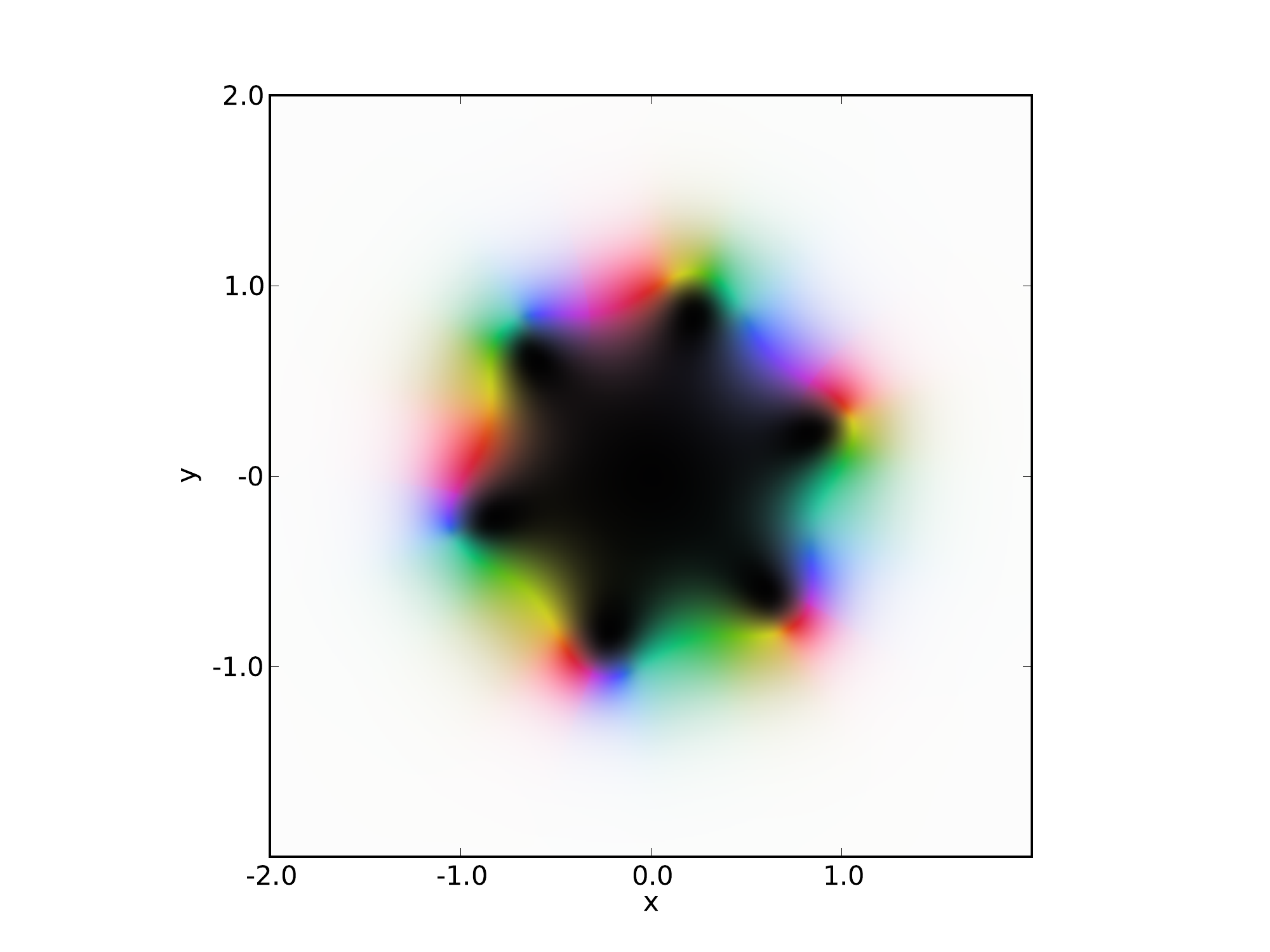}
\includegraphics[width=0.326\textwidth]{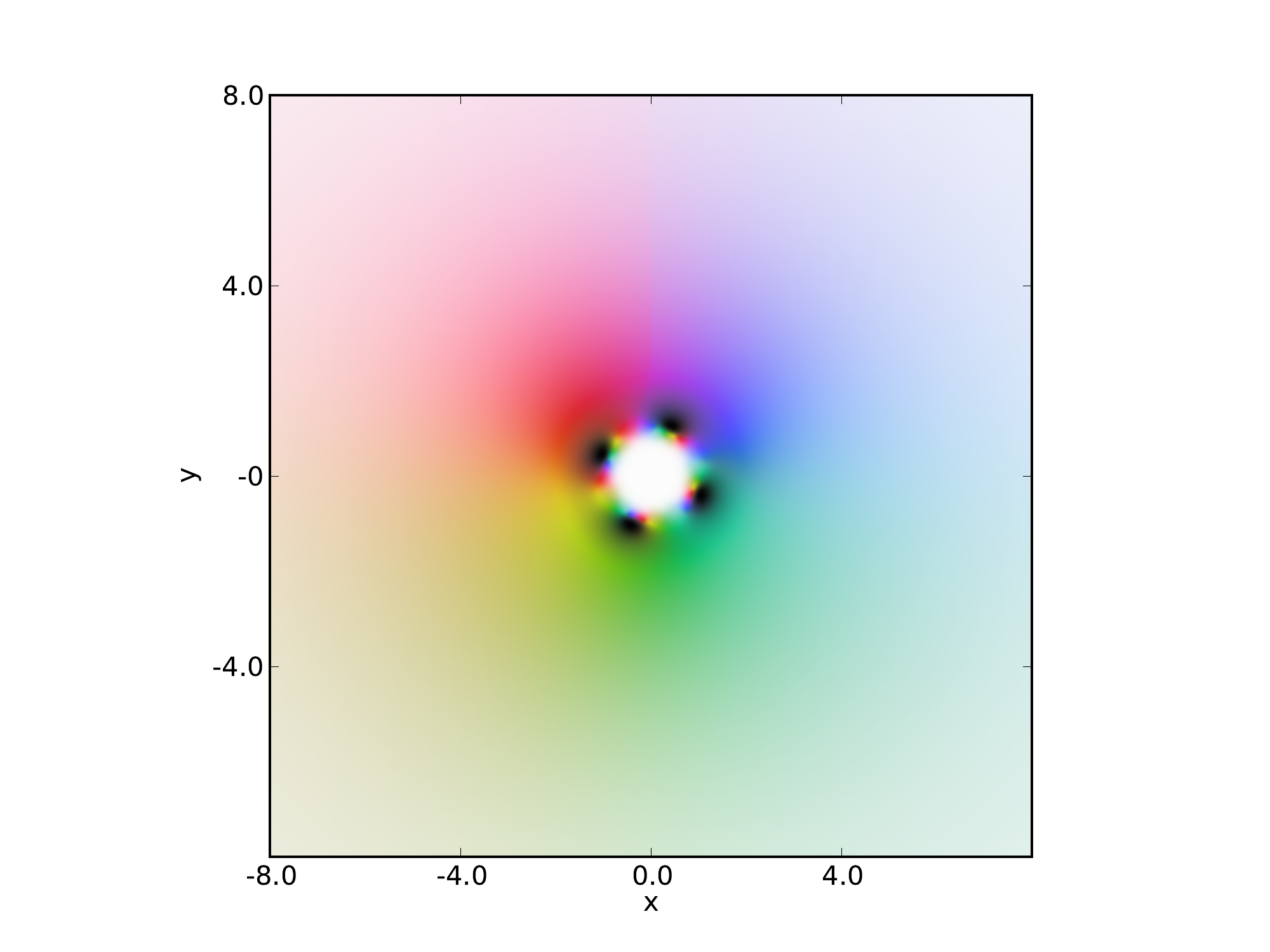}
   \vspace{0cm}
    \caption{{\scriptsize 
    Magnetic skyrmion solutions in the model with  DM term $(\bn,\nabla \times \bn)$ and the potential $\frac 12 (1-n_3)^2$. The value of the magnetisation vector $\bn$, assumed to be of unit length,   is shown in terms of the Runge colour sphere: the region near the south pole $n_3=-1$ is shown in black, and the `vacuum region' near the north pole $n_3=1$ is shown in white. Elsewhere, the longitudinal angle $\arctan (n_2/n_1)$  is mapped onto the colours red, green, blue, with intermediate colours interpolating. Solutions are determined by a choice of a holomorphic function $h$ and the formula \eqref{vh}. 
       Top from left to right:  the Bloch skyrmion with  $h(z)=0$,  the line defect with $h(z) = \frac i2 z$ and the anti-skyrmion with $h(z)=z$.  
     Bottom from left to right:  
    the `empty bag' with  $h(z)=2i/z$, the anti-skyrmion  of  charge $Q=5$ with  $h(z)=z^5$  and  the anti-skyrmion of charge $Q=4$ with  $h(z)=1/z^5 $.    }}
\label{piccies}
\end{figure}

This talk is about the critically coupled models recently proposed in \cite{BSRS} and \cite{Schroers}. These models require a particular  choice of  potential for any given DM term, but with this choice they can be viewed as a gauged version of the Belavin-Polyakov $O(3)$  sigma model \cite{BP}. In 
 particular, solutions can be obtained explicitly in terms of  holomorphic functions to the Riemann sphere $\C \cup \{\infty\}$.  In Fig.~\ref{piccies}, we show examples of such solutions  in a model with standard DM term $(\bn,\nabla \times \bn)$ and the potential $\frac 12 (1-n_3)^2$. They include the axisymmetric skyrmion configuration (which has topological charge $Q=-1$ in our conventions), a line defect ($Q=0$), an anti-skyrmion configuration ($Q=1$)  as well as bags and multi-(anti)-skyrmion configurations which show qualitative features  of the configurations studied numerically in \cite{RK} and \cite{FKATDS}.

 This  talk is designed  to explain the  models and the construction of their solutions from holomorphic data as  simply and directly as possibly. We sum up the  method of solution as a four-step recipe  in Sect.~3.    For details we refer the reader to the papers \cite{BSRS,Schroers}.

\section{Magnetic skyrmions and gauged sigma models}
\subsection{Formulating magnetic skyrme models as gauged sigma models}
\vspace{-0.1cm}
The most general  energy functional  for the  magnetisation field $\bn:\R^2 \rightarrow S^2$ which we consider has the  form 
\bee
\label{Skyrme1}
E[\bn]=\int_{\R^2}\left( \frac{1}{2}(\nabla \bn)^2 + \sum_{a=1}^3\sum_{i=1}^2 {\mathcal D}_{ai}(\partial_i  \bn  \times  \bn)_a +  V(\bn)  \right) d x_1d x_2,
\eee
where $V$ is a potential which may include  a Zeeman term and anisotropy terms, and $\mathcal D$ is the spiralization  tensor 
parametrising the DM interaction.

In the following we will use the complex coordinate $z= x_1+ix_2$   in the plane, and define associated derivatives in the standard way, so
$  \partial_z = \frac 12 (\partial_1 - i \partial_2)$ and 
$   \partial_{\bar z}  = \frac 12 (\partial_1 +i \partial_2)$. 
As shown in \cite{Schroers}, one can view the expression \eqref{Skyrme1} as the energy functional  of a gauged sigma model with a fixed $SU(2)$ background gauge field.  To see this,  it  is convenient to 
think of  an $SU(2)$  gauge field in the plane  simply  as a pair of vectors $\bA_1$ and $\bA_2$ in $\R^3$, one for each Cartesian direction in the plane,  which act on the magnetisation field $\bn\in \R^3$ via the vector product (the commutator of the $su(2)$ Lie algebra) so that the covariant derivative  and curvature are 
\bee
\label{covcur}
D_i\bn  = \partial_i\bn + \bA_i\times \bn, \quad i=1,2,  
\quad \bF_{12}= \partial_1\bA_2-\partial_2 \bA_1  + \frac 12 \bA_1\times \bA_2. 
\eee
The energy functional of the gauged non-linear sigma models studied in \cite{Schroers} is  
\bee
\label{localenergy}
E_A[\bn ]=  \int_{\R^2} \left( \frac 12 |D_1\bn|^2 +\frac 12  |D_2\bn|^2   - ( \bF_{12}, \bn) \right) dx_1  dx_2 .
\eee

In the application to magnetic skyrmions, the gauge field is fully determined by the  spiralization tensor: the Cartesian components  $\bA_1$ and $\bA_2$  are simply the negatives of the column vectors which make up the $3\times 2$ spiralization matrix $\mathcal D$. In symbols
\bee
\label{ADtranslation}
(\bA_i)_a = -  {\mathcal D}_{ai},
\eee
so that the DM  term can  be written  as 
\bee
\label{DMexp}
 \sum_{a=1}^3\sum_{i=1}^2 {\mathcal D}_{ai}(\partial_i \bn \times  \bn)_a =  \sum_{i=1}^2 (\bA_i\times  \bn,\partial_i \bn). 
 \eee
We  now need to pick a particular form of the  potential $V$  in \eqref{Skyrme1} to obtain a solvable model, namely
\bee
\label{specpot}
V_A(\bn) = \frac 12 |\bA_1\times \bn|^2 +  \frac 12 |\bA_2\times \bn|^2  -(\bn, \bA_1\times \bA_2).
\eee
With the choice  $V=V_A$, the magnetic skyrmion energy \eqref{Skyrme1}  equals the energy \eqref{localenergy}  of the gauged sigma model with the gauge field \eqref{ADtranslation}.  This is easily checked, and makes use of the fact that, for constant gauge fields,  one has $
\bF_{12} = \bA_1\times \bA_2$.  Note that the energy functional \eqref{localenergy} is invariant under $SU(2)$ gauge transformations, but that this gauge invariance is broken by the gauge choice \eqref{ADtranslation} to obtain the magnetic skyrme energy functional \eqref{Skyrme1} at critical coupling. The residual symmetries of the critically  model are discussed in \cite{BSRS} and \cite{Schroers}.

For a simple illustration, consider   
\bee
\label{simple}
\bA_1 =-\kappa \be_1, \bA_2= -\kappa \be_2, 
\eee  
where $\be_1 = (1,0,0)^t$ and $\be_2=(0,1,0)^t$ are the first two elements of the canonical frame for $\R^3$, and 
$\kappa>0$ is a real parameter. This  produces the standard DM term  $\kappa (\bn, \nabla\times \bn)$  and the potential
$V_A=\frac{ \kappa^2}{2}(1-n_3)^2$.   Expanding the square,  the potential is seen to be a particular linear combination of 
 an easy-plane anisotropy  potential with a Zeeman potential,  see \cite{BSRS}.  This model with $\kappa =1$ is the one whose solutions are shown in  Fig.~\ref{piccies}. 

\subsection{A Bogomol'nyi equation for  gauged sigma models}
Returning now to the case of a general gauge field, we use  various gauge-theoretical identities \cite{BSRS,Schroers}
to write the    energy  \eqref{localenergy} as 
\bee
 \label{Bogotrick}
 E_A[\bn]= \frac 1 2 \int_{\R^2} (D_1\bn +  \bn\times D_2\bn)^2 dx_1dx_2+  4\pi (Q + \Omega_A), 
\eee
where $Q$  is the integral  expression for the degree of $\bn$ 
\bee
\label{Qdef}
Q[\bn] =\frac {1}{4\pi} \int_{\R^2}( n, \partial_1 \bn \times \partial_2\bn) dx_1dx_2,
\eee
and 
$\Omega_A$ is a generalised version of what was called  total vortex strength in \cite{BSRS}:
\bee
\label{omdef}
\Omega_A[\bn]= -\frac {1}{4\pi} \int_{\R^2} (\partial _1(\bA_2, \bn) -\partial_2(\bA_1, \bn)) dx_1dx_2.
\eee
If these integrals are well-defined, they only depend on global properties of $\bn$ and on its boundary behaviour. If the latter is kept fixed, the energy is therefore minimised when the  square in \eqref{Bogotrick} vanishes, i.e. when  the Bogomol'nyi equation holds. This is a gauged version of the Bogomol'nyi equations in the standard Belavin-Polyakov model \cite{BP}, but with a definite sign:
\bee
\label{Bogon}
 \bn \times D_1 \bn = D_2\bn.
\eee
Ths Bogomol'nyi equation implies the variational equation of the energy functional \eqref{localenergy}, see \cite{Schroers}. 

In the context of magnetic skyrmions, the equation \eqref{Bogon} first appeared in \cite{DM} where it was noticed that, for a certain family of potentials $V$,  it characterises  the energy minimisers in the $Q=-1$ sector of the theory with the standard DM  term $(\bn,\nabla\times \bn)$.  The role of this equation  in critically coupled magnetic skyrme models for arbitrary degree $Q\geq -1$   was observed and explored in \cite{BSRS}.  Its role and solvability  in the more general  gauged sigma model and  the associated magnetic skyrmion models is the subject of \cite{Schroers}. Generalised versions of this equation have been studied  in differential geometry as vortex equations for maps from Riemann surfaces into K\"ahler manifolds which permit  the action of  a Lie group, but in that case the gauge field typically obeys  a second, coupled equation \cite{CGS}. The relation between these vortex equations  and the  equation \eqref{Bogon} with a fixed background as proposed in  \cite{Schroers} was clarified in \cite{Walton}.

\subsection{Boundary contributions to the energy}
\label{boundary}
As far we are aware it is not known for which class of magnetisation fields $\bn$ the general energy functional \eqref{Skyrme1}  is well-defined and finite. For the standard DM term $(\bn, \nabla \times \bn)$ and a certain class of potentials $V$, this question is answered  in  \cite{Melcher} and \cite{DM}, where it was  also pointed that, for  analytical reasons, it is preferable to modify the energy functional by adding the   boundary term
\bee
\label{kappabound}
E_{\kappa ,\infty}[\bn]= -\kappa \int_{\R^2}  (\partial _1 n_2 -\partial_2 n_1) dx_1dx_2.
\eee
Adding this  term to the energy effectively modifies  the DM term:   $\kappa(\bn,\nabla\times \bn) $  is replaced by  $ \kappa((\bn-\be_3),\nabla\times \bn)$ where $\be_3=(0,0,1)^t$, and this  is the term  considered in \cite{Melcher,DM}.

In  the context of gauged sigma models, it was proposed  in \cite{Schroers} that one should more  generally add the boundary term
\bee
\label{Abound}
E_{A,\infty} [\bn] = -4\pi \Omega_A[\bn] = \int_{\R^2} (\partial _1(\bA_2, \bn) -\partial_2(\bA_1, \bn)) dx_1dx_2 
\eee
to obtain a well-defined variational problem.  Clearly, \eqref{Abound}  reduces to \eqref{kappabound} for  the simple gauge field  \eqref{simple}. 

Adding the term  \eqref{Abound} to  the energy \eqref{localenergy}  has a number of advantages, at least from an analytical point of view. It does not change the Euler-Lagrange equation one obtains for  variations which vanish rapidly at infinity, but its inclusion means that one can allow for variation  with a slower fall-off.  We refer the reader to \cite{Schroers} for details. Furthermore, the  study of solutions of arbitrary degree in \cite{BSRS}  shows that the modified energy is  well-defined   for some solutions for which  the unmodified energy integral  \eqref{localenergy} is not.

Geometrically, the unmodified energy \eqref{localenergy} has a natural interpretation when evaluated on a solution of the Bogomol'nyi equation  as the equivariant degree of that solution \cite{Walton}. By contrast,  the 
 modified energy  evaluated on  a  solution $\bn$  of the Bogomol'nyi equation is equal to the integral expression for the degree:
 \bee
 \label{moden}
E_A[\bn]+ E_{A,\infty}[\bn]= 4\pi Q[\bn] \quad \text{if} \quad  \bn \times D_1 \bn = D_2\bn.
 \eee

\section{Exact magnetic skyrmions}

\subsection{The general solution in four easy steps}
Since the magnetic skyrmion energy functional  \eqref{Skyrme1} with the potential \eqref{specpot} is a particular example of the energy for a gauged sigma model of the form \eqref{localenergy}, we can obtain an infinite family of solutions of the variational equations  by solving \eqref{Bogon}. Here we focus on the formula needed for  magnetic skyrmions, so for constant gauge fields. In that case, the solution of \eqref{Bogon} can be obtained via the following recipe. For details we again refer to  \cite{Schroers}.

\begin{enumerate}
\item[(I)] {\em Complex coordinate for the magnetisation:} \;
In order to write down the solution, one needs to work in terms of a complex stereographic coordinate for the magnetisation field $\bn$. It is  given by stereographic projection from the south pole, or algebraically by 
\bee
\label{wn}
w =  \frac{n_1+in_2}{1+n_3}.
\eee
\item[(II)] {\em Complexified gauge field:}\;
 Next, one needs to write the gauge field explicitly as an $su(2)$ matrix-valued gauge field on $\R^2$  according to 
 \bee
 A_i=\sum_{a=1}^3 A_i^a t_a, \quad i=1,2, 
 \eee
  where  $t_a=-\frac i2 \tau_a$, and $\tau_a$ are the Pauli matrices. In fact we require the complex linear combination
   \bee
A_{\bar z} = \frac 12 (A_1+iA_2). 
 \eee   
   In the case at hand, this is a constant, complex and traceless $2\times 2$ matrix, so generically  an element of $sl(2,\C)$.
   \item[(III)] {\em Solution in complex coordinates:}\;
The solution of the Bogomol'nyi equation is given in terms of the exponential
 \bee
 \label{gmatrix}
g(\bar z)= \exp(- \frac 12 (A_1+iA_2)\bar z)=\begin{pmatrix} a(\bar z)  & b(\bar z)  \\ c(\bar z) & d(\bar z)  \end{pmatrix}, 
 \eee
which  is a  $2\times 2$ matrix function of $\bar z$ with determinant one. 
The general solution of \eqref{Bogon} in stereographic coordinates is  
 \bee
\label{mastersol}
w(z,\bar z) = \frac{c(\bar z)  + d(\bar z)  f(z)  }{a(\bar z)  +b(\bar z) f(z) },  
\eee
where $f$ is an arbitrary  holomorphic  function from $\C$ into $\CP^1\simeq \C \cup \{\infty\}$ (in particular it  is allowed to take the value $\infty$). 
\item[(IV)] {\em Translating back into Cartesian coordinates:}\;
Substitution of the general solution \eqref{mastersol} into  the   inverse of \eqref{wn}
\bee 
\label{nw}
 n_1+in_2= \frac{ 2 w }{1+|w|^2}, \quad n_3= \frac{1-|w|^2}{1+|w|^2},
 \eee
 yields an explicit (but possibly complicated) formula for the  magnetisation field. 
 \end{enumerate}
The energy density of Bogomol'nyi solutions is either the degree density or  the sum of the degree density and the vorticity, depending on the choice of energy functional, see our discussion in Sect.~\ref{boundary}. Expressions for both directly in terms the stereographic coordinates are given in \cite{BSRS,Schroers}.

\subsection{Examples}
\label{exsect}
\noindent {\em Axisymmetric DM terms:}\;
As discussed in \cite{Schroers}, the DM term is invariant under rotations in the plane and  simultaneous  rotations of the magnetisation field about a suitable axis if and only if $\bA_1$ and $\bA_2$ are orthogonal and have the same length. In that case $(  \bA_1, \bA_2, \bA_1\times \bA_2)$ is an oriented and (up to scaling) orthonormal basis of $\R^3$.  With $|\bA_1|=|\bA_2|=\kappa$, the potential for the solvable model is  conveniently expressed in terms of $\hA:=\bA_1\times\bA_2/\kappa^2$ as 
 \bee
 V_A(n) = \frac {\kappa^2}{2}   \left(1 - (\bn,\hA)\right)^2 = \frac{\kappa^2}{8}(\bn-\hA)^4.
 \eee
 The DM term \eqref{DMexp}  and the integrand of the boundary term \eqref{Abound} combine neatly into 
 $ \sum_{a=1}^3\sum_{i=1}^2{\mathcal D}_{ai}(\partial_i \bn \times  (\bn-\hA))_a $ in this case.
 For the simple case  \eqref{simple} with  DM term  $\kappa (\bn, \nabla\times \bn)$  and  potential  
$V(\bn)=\frac{ \kappa^2}{2}(1-n_3)^2$, one checks that  the matrix \eqref{gmatrix} is
\bee
\label{rhoskyr}
g  =\begin{pmatrix}
1 & - \frac{ i }{2} \kappa  \bar z \\ 0 & \phantom{-}1
\end{pmatrix}.
\eee
The solution \eqref{mastersol} is best written in terms of the inverse coordinate 
 $v=1/w$ as
\bee
\label{vh} 
v= - \frac{ i }{2} \kappa \bar z +h,
\eee
where $h = 1/f$ is, like $f$, an arbitrary holomorphic map $\C \rightarrow \CP^1$. The simplest choice $h=0$ leads to the Bloch hedgehog skyrmion   with $Q=-1$ and  the Belavin-Polyakov profile function $ 
\theta(r) = 2 \arctan \left(\frac{r}{2\kappa}\right)$, as already  noticed in \cite{DM}. Many other solutions are discussed in \cite{BSRS}, and our Fig.~\ref{piccies} shows the solutions one obtains for different choices of rational functions  $h$.   It follows from the calculations in \cite{BSRS} that the degree  of a  skyrmion configuration  depends on the parameter
 $L =  \lim_{|z|\rightarrow \infty} 
\left |(2h)/(\kappa \bar z) \right|$. For configurations determined by \eqref{vh} with rational $h(z)=p(z)/q(z)$, where   $p$ and $q$ are polynomials of degree $M$ and $N$, it  is 
\bee
Q [\bn]=\begin{cases}  M  & \text{if} \quad L  >1 \\
 N & \text{if} \quad  L =1 \\
N-1& \text{if} \quad  L <1.
\end{cases}
\eee
This shows in particular that in this model there are infinitely many solutions of the Bogomol'nyi equation \eqref{Bogon} for each integer degree $Q\geq -1$. The modified energy \eqref{moden} takes the values $4\pi Q$ on these solutions. 

\noindent {\em Rank one DM interactions:} \;
 The spiralization tensor has rank one when  $\bA_1$ and $\bA_2$ are linearly dependent, so $\bA_1 \times \bA_2=0$. 
 In this case,   the curvature $\bF_{12}$  vanishes and the gauge field (and therefore the DM interaction) can be removed by an $SU(2)$  gauge transformation. The solvable model with the potential \eqref{specpot}  can therefore be mapped into the standard
  Belavin-Polyakov  $O(3)$ sigma model \cite{BP}.    
   It follows immediately   that solutions of \eqref{Bogon} exist for any integer degree $Q$  in these models, and that their  energy \eqref{moden} is $4\pi |Q|$. In particular it follows that 
   skyrmions  and  anti-skyrmions of equal and opposite degree have  the same  energy.  This result was derived in \cite{genDMI}  for $|Q|=1$ in more general  rank one models.
Example solutions of  solvable rank one  models  and their properties are discussed in  \cite{Schroers} and also \cite{HFMNS}. We note that a   similar  reformulation in terms of a flat gauge field was recently applied to a rather different ferromagnetic model in \cite{SW}.

  \section{Conclusion}
We have shown that solvable  models of magnetic skyrmions exist for any DM interaction term. Even though they require fine-tuning of the potential, their exact  solutions shed light on  qualitative properties of solutions in more general models.  These includes general features of multi-(anti)-skyrmion configurations  such as  the  appearance of  $Q+1$ maxima in the energy density   of certain  charge $Q>0$ configurations (as shown in  the $Q=5$ solution  in Fig.~\ref{piccies}), or the deformation of a skyrmion to an anti-skyrmion via a line defect,  as shown in the top row of  Fig.~\ref{piccies}  and discussed in some detail in \cite{BSRS}.

 The solvable models  also shed light on  the crucial influence of the DM interaction on the relative energy of skyrmions  compared  to anti-skyrmions.  Our short discussion  illustrates the more general findings of \cite{genDMI}. In the axisymmetric models  of Sect.~\ref{exsect},  $Q=-1$  skyrmions have energy $-4\pi$ whereas  $Q=1$ anti-skyrmions have the opposite energy $4\pi$ (this can be reversed by a different choice of solvable model, see \cite{BSRS}). In rank one models, by contrast, skyrmions and anti-skyrmions have the same energy. It was shown in \cite{genDMI} that  models with generic spiralization tensors should interpolate between these two extremes, and it would be interesting to explore this in the solvable models with generic DM terms.

To end,  we note that the  language of gauged non-linear   sigma models provides  a rare and rather beautiful link between pure mathematics and real physics by connecting  the geometry of holomorphic maps and  vortex equations as discussed in \cite{CGS} with  magnetic skyrmions.  In fact,  simply allowing the gauge field to depend non-trivially on space may provide further applications, for example to  to the study of impurities as discussed in  \cite{AQW} and \cite{Schroers}.

\paragraph{Acknowledgements}
I thank Bruno Barton-Singer for sharing his Python code for generating the plots in Fig.~\ref{piccies},  Calum Ross for pointing out an  error in an earlier version of this paper, and the referee for constructive comments. 
%
%


\begin{thebibliography}{100}
%


\bibitem{BY}
A.~N.~Bogdanov,  D.~A.~Yablonskii,  Thermodynamically Stable `Vortices' in Magnetically  Ordered  Crystals.  The  Mixed  State  of Magnets,  Zh.~Eksp.~Teor.~Fiz.  {\bf 95},  178 (1989) .



\bibitem{NT}
N.~Nagaosa,  Y.~Tokura, Topological Properties and Dynamics of Magnetic Skyrmions, Nature Nanotechnology {\bf 8},    899 (2013), \doi{10.1038/NNANO.2013.243}.



\bibitem{Dzyaloshinskii} I.~Dzyaloshinskii, A Thermodynamic Theory of `Weak' Ferromagnetism of Antiferromagnetics, J.~Phys.~Chem.~Solids {\bf 4},   241 (1958),  \doi{10.1016/0022-3697(58)90076-3}. 

\bibitem{Moriya} T.~Moriya,  Anisotropic Superexchange Interaction and Weak Ferromagnetism,  Phys.~Rev.~{\bf 120},  91 (1960), \doi{10.1103/PhysRev.120.91}.


\bibitem{FCS}
A.~Fert, V.~ Cros, J.~ Sampaio, Skyrmions on the Track,  Nature Nanotechnology {\bf  8},
152 (2013),  \doi{10.1038/nnano.2013.29}.

\bibitem{Melcher}
C.~Melcher,  Chiral Skyrmions in the plane, Proc.~R.~Soc.~A {\bf 470},  20140394 (2014), \doi{10.1098/rspa.2014.0394}.


\bibitem{DM} L.~D\"oring and C.~Melcher, Compactness Results for Static and Dynamic Chiral Skyrmions near the Conformal Limit, Calc.~Var. {\bf 56}, 60 (2017), \doi{10.1007/s00526-017-1172}.



\bibitem{genDMI}
M.~Hoffmann, B.~Zimmermann, G.~P.~ M\"uller, D.~Sch\"urhoff, N.~S.~Kiselev, C.~Melcher, S.~Bl\"ugel,  
Antiskyrmions Stabilized at Interfaces by Anisotropic Dzyaloshinskii-Moriya Interactions,  Nature Communications {\bf  8},  308 (2017), \doi{10.1038/s41467-017-00313-0}. 



\bibitem{BSRS} B.~Barton-Singer, C.~Ross,  B.~J.~Schroers, Magnetic Skyrmions at Critical Coupling. Commun. Math. Phys. (2020) \doi{10.1007/s00220-019-03676-1}; \url{https://arxiv.org/abs/1812.07268}. 


\bibitem{Schroers} B.~J.~Schroers, Gauged Sigma Models and Magnetic Skyrmions, SciPost Phys. {\bf 7}, 030 (2019) \doi{10.21468/SciPostPhys.7.3.030}.


\bibitem{BP} A.~A.~Belavin, A.~M.~Polyakov, Metastable States of Two-Dimensional Isotropic Ferromagnets, JETP Letters  {\bf 22}, 245 (1975).

\bibitem{RK}
F.~N.~Rybakov and N.~S.~Kiselev, 
Chiral Magnetic Skyrmions with Arbitrary Topological Charge (`Skyrmionic Sacks'),  Phys. Rev. B {\bf 99}, 064437 (2019), \doi{10.1103/PhysRevB.99.064437}.



\bibitem{FKATDS}
D.~Foster, C.~Kind, P.~J.~Ackerman, J.~S.~B.~Tai, M.~R.~Dennis and I.~I.~Smalyukh,
Two-dimensional Skyrmion Bags in Liquid Crystals and Ferromagnets, Nature Physics {\bf 15},  655 (2019), \doi{10.1038/s41567-019-0476-x}.



\bibitem{CGS} K.~Cieliebak, A.~R.~Gaio and D.~A.~Salamon, J-holomorphic Curves, Moment Maps, and Invariants of Hamiltonian Group Actions, Internat. Math. Res. Notices {\bf 10 } 831 (2000), \doi{10.1155/S1073792800000453}.


\bibitem{Walton}
E.~Walton, Some exact Skyrmion solutions on curved thin films, \url{https://arxiv.org/abs/1908.08428}.



\bibitem{HFMNS}
M.~Hongo, T.~Fujimori, T.~Misumi, M.~Nitta, N.~Sakai, Instantons in Chiral Magnets, \url{https://arxiv.org/abs/1907.02062}.

\bibitem{SW} M.~Speight and T.~Winyard, Skyrmions and Spin Waves in Frustrated Ferromagnets at Low Applied Magnetic Field, \url{https://arxiv.org/abs/1909.07970}.

\bibitem{AQW}C.~Adam,  J.~M.~Queiruga, 
A.~Wereszczynski,  BPS Soliton-impurity Models and Supersymmetry,  J. High Energy Phys. 2019: 164 (2019),
\doi{10.1007/JHEP07(2019)164}.





\end{thebibliography}
\end{document}